\documentclass[lettersize,journal]{IEEEtran}
\usepackage{amsmath,amsfonts,amssymb}
\usepackage{array}
\usepackage{booktabs}
\usepackage{textcomp}
\usepackage{stfloats}
\usepackage{url}
\usepackage{verbatim}
\usepackage{graphicx}
\usepackage{cite}
\usepackage{xcolor}
\AtBeginDocument{%
  }

\begin{document}

\title{FillGauss: Fine-Grained Filling-Aware Impact Sound Generation for 3D Gaussian Splatting}

\author{Chen Yang, Ganye Wen, Bin Huang, Jiayi Lyu, Zehai Niu, Linlin Shen, Jinbao Wang}

\maketitle

\begin{abstract}
Synthesizing physically plausible impact sounds from visual observations remains a great challenge in multi-modal AI. Existing 3D-aware audio generation methods primarily model the surface geometry of hollow rigid bodies. However, they fundamentally overlook internal filling states, a critical physical factor that drastically modulates acoustic resonance and damping. To address this issue, we have defined a new task called Fine-Grained Filling-Aware Impact Sound Generation. As a foundational step, we first introduce the fine-grained fill-aware dataset (FillImpact), a pioneering multi-modal collection comprising over 5,000 rigorous acoustic recordings from 88 diverse real-world objects. It captures impact interactions with varying internal contents (i.e., water, rice), a continuous range of fill levels, and distinct striker materials. Furthermore, comprehensive acoustic analysis confirms that the collected data closely aligns with established physical laws governing acoustic resonance and damping, indicating its suitability for physically grounded modeling. Building on this dataset, we propose a novel generative framework (FillGauss) that integrates 3D Gaussian Splatting (3DGS) with internal state conditioning for sound generation. By fusing 3DGS geometric features, precise 3D spatial strike coordinates, and fine-grained textual physical conditions within a latent diffusion architecture, FillGauss enables position-aware, striker-aware, and filling-aware audio generation. Extensive experiments demonstrate that our approach could generate high-fidelity impact sounds that adhere to underlying physical principles, establishing a new state-of-the-art for physically grounded cross-modal audio generation.
\end{abstract}

\begin{IEEEkeywords}
Impact Sound Generation, 3D Gaussian Splatting, Cross-Modal Synthesis, Latent Diffusion Models.
\end{IEEEkeywords}

\begin{figure*}[t]
  \centering
  \includegraphics[width=\textwidth]{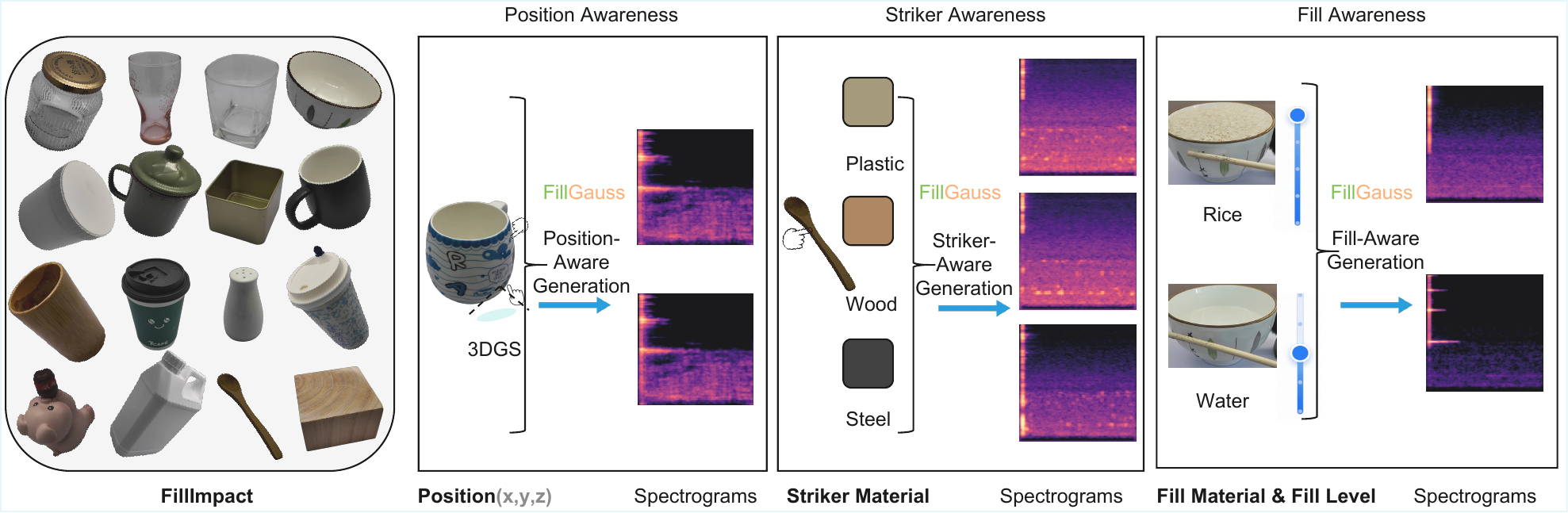}
  \caption{Overview of the FillGauss framework capabilities. Left: Some objects of FillImpact. Right: Given a 3D Gaussian Splatting representation, our model (FillGauss) enables fine-grained impact sound generation. FillGauss successfully models complex physical interactions to generate physically plausible acoustic spectrograms (audios) across three distinct dimensions: position-aware, striker-aware, and fill-aware.}
  \label{fig:pipeline_1}
\end{figure*}

\section{Introduction}

The physical interaction between objects, such as striking or tapping, produces rich acoustic signals that provide vital clues about an object's material, shape, and internal state. In immersive environments like Virtual Reality, interactive gaming, and robotic manipulation~\cite{thankaraj2023sounds,taylor2009resound,owens2016visually}, synthesizing physically plausible impact sounds is crucial for multi-sensory realism and scene understanding. In recent years, multi-modal learning has made significant strides in audio-visual generation. However, generating physically accurate impact sounds directly from 3D visual representations remains a formidable open challenge. Most current methodologies rely on 2D videos or 3D meshes, implicitly assuming that objects are strictly hollow or structurally uniform. They fundamentally fail to account for the internal filling state of an object, which drastically alters its acoustic resonance.

To address this fundamental limitation, we formally define a novel task, which is \textit{Fine-Grained Filling-Aware Impact Sound Generation}. To support our new task, we introduce \textbf{FillImpact} dataset. This dataset meticulously captures the impact audio of containers under varying internal states, comprising over 5,000 rigorous acoustic recordings across 88 diverse real-world objects from six external material categories. It systematically records interactions involving different fill materials (e.g., water and rice), continuous fill levels, and distinct striker materials (plastic, wood, and steel). Our in-depth dataset analysis verifies that the recorded acoustic behaviors align perfectly with established physical principles. For instance, the acoustic decay coefficient ($\alpha$) varies significantly between granular and liquid media, and the normalized dominant frequency exhibits a predictable, inverse relationship with the fill level, consistently obeying the classic harmonic oscillator model $f \propto \sqrt{k/m}$. Furthermore, the spectral centroid is naturally influenced by the hardness of the striker material.

\begin{figure}[th]
  \centering
  \includegraphics[width=1\linewidth]{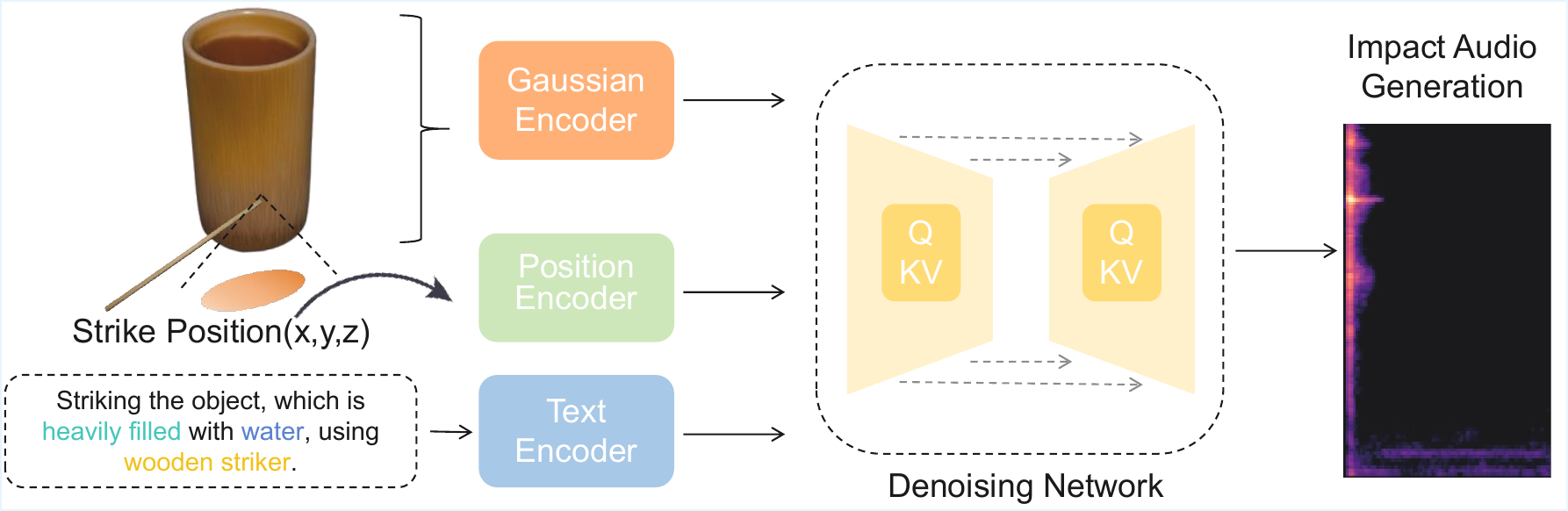}
  \caption{Simplified architecture of the FillGauss framework. It illustrates how the Gaussian, Position, and Text Encoders extract multi-modal features to guide the Denoising Network for fine-grained, filling-aware impact sound generation.}
  \label{fig:pipeline_2}
\end{figure}

Building upon these robust physical priors, we propose FillGauss, a novel generative framework designed for fine-grained, filling-aware impact sound generation. As illustrated in Figure~\ref{fig:pipeline_2}, FillGauss is the first model to seamlessly integrate 3D Gaussian Splatting (3DGS) with internal state conditioning. The architecture employs a Gaussian Encoder to capture the rich 3D surface representation, alongside a Position Encoder to extract the exact 3D strike coordinates. To effectively guide the physical generation process, we utilize a Text Encoder that processes detailed condition descriptions (e.g., "The object is hit with a wooden striker. It is heavily filled with water."). These multi-modal features are deeply entangled using a self-attention module and a cross-attention mechanism, which subsequently condition a pre-trained latent diffusion Denoising Network. Extensive experiments demonstrate that FillGauss successfully tackles this novel task, achieving state-of-the-art acoustic fidelity in precise position-aware, striker-aware, and fill-aware sound generation.

In summary, the main contributions of this paper are:
\begin{itemize}
    \item We formally define a novel task, called Fine-Grained Filling-Aware Impact Sound Generation. This task enhances interactive experiences with 3DGS-reconstructed objects by enabling filling-aware auditory feedback.
    
    \item We construct FillImpact, a fine-grained filling-aware dataset comprising over 5,000 impact recordings from 88 3D objects. Rigorous acoustic analysis demonstrates that the collected data closely aligns with established physical laws.
    
    \item We propose FillGauss, a flexible framework that integrates 3D Gaussian Splatting (3DGS) with internal filling conditions to enable physically aware sound generation. 
    
    \item Comprehensive evaluations show that FillGauss achieves state-of-the-art performance, consistently generating realistic, physically plausible acoustic spectrograms. These results accurately capture variations across strike position, striker material, and internal filling states.
    
\end{itemize}

\section{Related Work}

\subsection{Impact Sound Generation}

\textbf{General Impact Sound Generation.} The synthesis of impact sounds has been extensively explored, historically transitioning from physics-based simulations to modern data-driven approaches. 
Traditional physics-based methods rely on modal analysis or finite element methods (FEM) to simulate object vibrations by solving wave equations~\cite{van2003modal,ren2013example,fang2024dynamic,bing2024dynamic}. 
While these methods yield physically accurate acoustics, they demand precise material parameters (e.g., Young's modulus, damping coefficients) and incur prohibitive computational costs, limiting their scalability in interactive scenarios.
To overcome these computational bottlenecks, recent advances in deep learning have catalyzed a shift toward cross-modal data-driven generation. These approaches learn complex mappings from visual observations to auditory responses, utilizing a variety of input modalities ranging from 2D videos~\cite{owens2016visually,zhou2018visual,chen2024action2sound} to basic 3D representations~\cite{liu2024sonicsense,su2023physics}. Furthermore, hybrid frameworks like DiffSound~\cite{yang2023diffsound} attempt to integrate implicit shape representations with diffusion models to enhance acoustic realism. 
However, a fundamental limitation persists across most existing data-driven and hybrid methodologies, they predominantly treat objects as homogeneous, rigid shells. By primarily focusing on external geometry or 2D visual cues, the complex acoustic modulation caused by internal fillings remains largely unexplored.

\textbf{3D-aware Impact Sound Generation.} To achieve spatially accurate audio synthesis, the paradigm has increasingly shifted towards leveraging explicit 3D reconstructions rather than 2D visual inputs~\cite{luo2022learning,brunetto2024neraf,chen2020soundspaces,li2022neuralacoustic}. 
By encoding the spatial structure of an object or environment, 3D-aware methods provide a robust foundation for simulating position-dependent acoustic responses. Among modern 3D representations, 3D Gaussian Splatting (3DGS)~\cite{kerbl20233dgs} has emerged as a highly effective medium due to its explicit geometric capture and high-fidelity texture rendering. 
Consequently, recent research has begun integrating 3DGS with generative audio models. 
Notably, SonicGauss~\cite{wang2025sonicgauss} utilizes a diffusion-based architecture~\cite{hung2024tangoflux}, replacing the traditional text encoder with a 3DGS encoder to achieve position-conditioned impact sound generation directly from the reconstructed 3D surface. 
Nevertheless, because 3DGS is inherently optimized for surface-level novel view synthesis, it explicitly captures the outer geometric hull but remains completely oblivious to the internal physical state of the object. 
This fundamental structural blind spot restricts the physical plausibility of current 3D-aware models when interacting with containers holding liquids or granular materials, directly motivating our filling-aware FillGauss architecture.
\begin{figure*}
  \centering
  \includegraphics[width=0.85\textwidth]{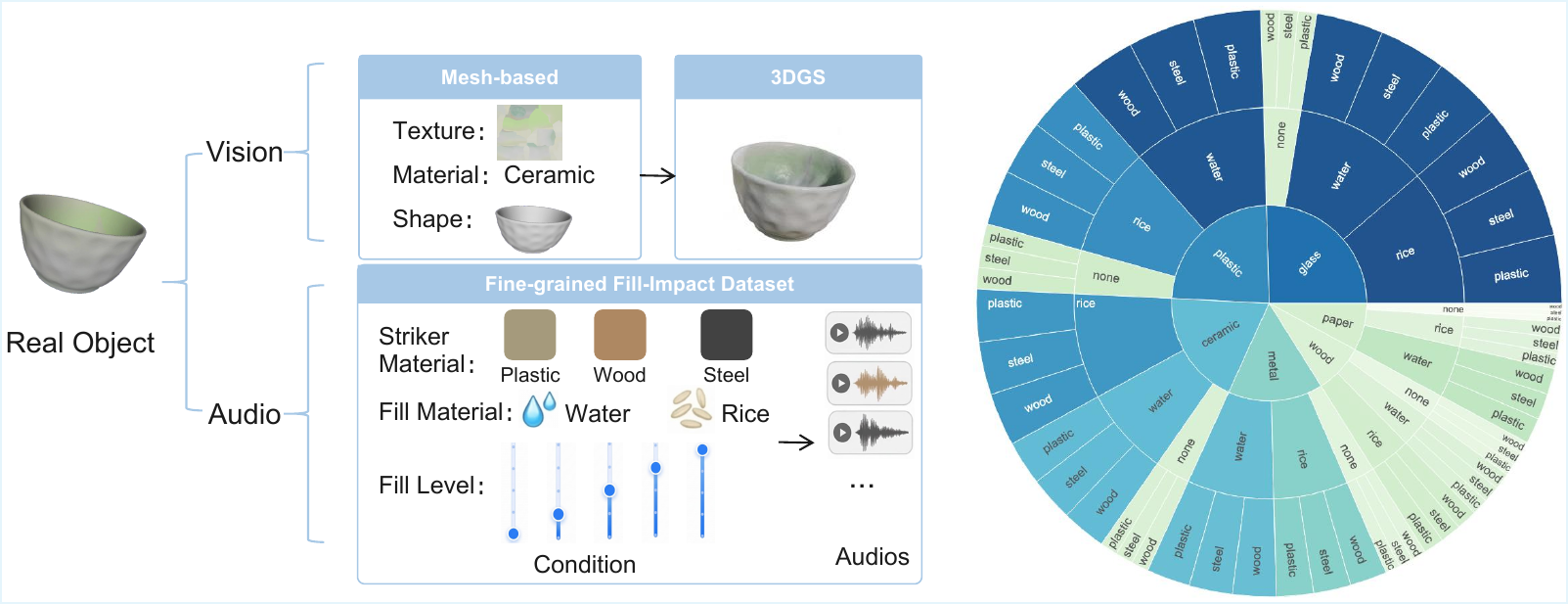} 
  \caption{Overview of FillImpact. Left: The construction of FillImpact. Right: Hierarchical distribution of FillImpact. The sunburst chart details the compositional breakdown from the inner to outer rings: outer container materials (e.g., glass, plastic, ceramic), internal fill states (water, rice, or hollow), and the interacting striker materials (steel, wood, plastic). The proportional area of each segment represents the relative number of instances in that specific configuration.}
  \label{fig:dataset_overview}
\end{figure*}
\subsection{Acoustic Effects of Internal Fillings}
Internal materials inside containers can substantially influence impact sounds~\cite{rayleigh1896theory}. 
When a container is struck, the resulting acoustic response depends on the coupled dynamics between the container structure and the internal material. The internal state has two directions that affect sound generation.
First, For liquid-filled containers, increasing the liquid level introduces an \textit{added mass effect}, which effectively increases the vibrating mass of the container walls and lowers structural resonance frequencies~\cite{rayleigh1945theory,kinsler2000fundamentals}. 
At the same time, the air cavity inside the container can exhibit Helmholtz resonance~\cite{webster2010use}, where the resonance frequency depends on the remaining air volume~\cite{fletcher1998physics}. 
These interacting effects lead to complex changes in acoustic characteristics as the filling level varies.
Second, Granular materials exhibit even more complex behavior due to frictional contacts, particle collisions, and force-chain dynamics~\cite{jaeger1996granular,makse2004granular}. 
Impact excitation in granular media typically produces strongly damped responses with dominant low-frequency components and short reverberation times. 
Because these heterogeneous physical behaviors are visually occluded from the outside, relying solely on external surface geometry is fundamentally insufficient. This necessitates a framework that explicitly entangles geometric representations with internal physical conditions.

\subsection{Impact Sound Datasets}
High-quality datasets are crucial for training and evaluating multimodal sound synthesis models. 
Existing impact sound datasets generally fall into two categories: simulation-based datasets and real-world recordings. Simulation-based datasets rely on physics engines or finite element analysis to generate impact sounds under controlled conditions. 
For example, Sound-20K~\cite{zhou2018sound20k} contains thousands of synthetic recordings of object impacts generated in virtual environments. 
More advanced datasets such as ObjectFolder~\cite{gao2021objectfolder} and ObjectFolder~2.0~\cite{gao2022objectfolder2} provide multimodal object representations and enable querying sound responses at arbitrary surface locations. 
However, these datasets primarily focus on rigid objects and fundamentally lack the modeling of internal material variations. Real-world datasets attempt to reduce the simulation-to-real gap by capturing physical interactions. 
The Greatest Hits dataset~\cite{owens2016visually} collects large-scale audio-video recordings of humans striking various materials in natural environments. 
More controlled datasets such as RealImpact~\cite{li2023realimpact} employ automated mechanical systems to produce repeatable impacts and record spatial acoustic responses using dense microphone arrays. 
While these datasets improve realism and spatial coverage, they predominantly feature solid or empty objects, lacking the fine-grained annotations of internal filling states required to learn the aforementioned complex acoustic dynamics.


\section{FillImpact}

We introduce FillImpact, a comprehensive multimodal collection comprising 88 real-world objects across various material categories. As illustrated in Figure \ref{fig:dataset_overview} Left, each instance in our dataset is rigorously paired with visual representations and acoustic interaction responses. The statistical distribution of our dataset is presented in Figure \ref{fig:dataset_overview} Right, which comprehensively illustrates the hierarchical composition of container materials, internal fill states, and striker types. Additional dataset statistics are provided in Appendix. In the following subsections, we detail our pipeline across three distinct dimensions: Visual Representation, Acoustic Acquisition, and Acoustic Analysis.

\begin{figure*}
  \centering
  \includegraphics[width=1\textwidth]{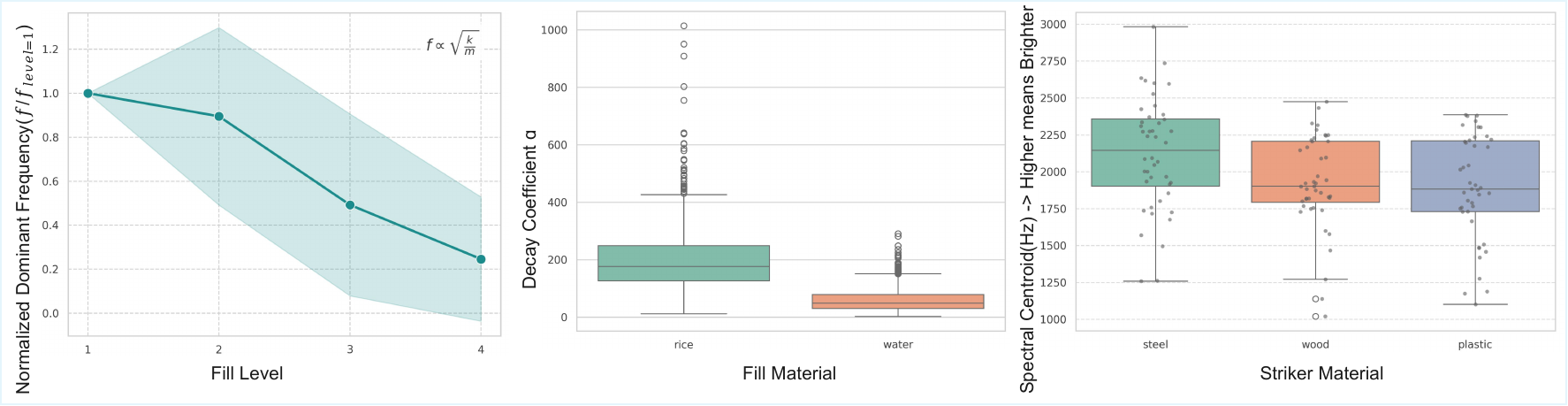}
  \caption{Acoustic analysis of FillImpact. Left: The normalized dominant frequency decreases as the fill level increases, strictly following the physical principle $f\propto\sqrt{k/m}$. Middle: The decay coefficient ($\alpha$) exhibits distinct distributions for different fill materials, demonstrating clear acoustic discriminability between granular (rice) and liquid (water) states. Right: The spectral centroid shows significant variation across striker materials, indicating that acoustic brightness effectively distinguishes the hardness and physical properties of the striker.}
  \label{fig:dataset_analy}
\end{figure*}

\textbf{Visual Representation.} To construct accurate geometric priors, we capture high-fidelity surface meshes using a Revo Scan 3D scanner. Seamless texture maps are then generated via Hunyuan 3D 2.1~\cite{li2024hunyuan3d} using single-view reference images extracted from 360-degree rotational videos captured under uniform lighting. These geometric and textural assets are then unified into the standard glTF format~\cite{gltf20}. 
For 3D Gaussian Splatting (3DGS)~\cite{kerbl20233dgs} optimization, we render 30 uniformly distributed multi-view images from the glTF models. During rendering, virtual lighting is strictly aligned with the camera optical axis to maximize fidelity, and precise camera extrinsics are concurrently exported. To avoid local minima, 3DGS training is spatially initialized using point clouds extracted from the glTF models and optimized for 10,000 iterations. Finally, the trained 3D Gaussians are normalized to ensure compatibility with the Point Transformer architecture~\cite{zhao2021point, chen2024splatformer}.

\textbf{Acoustic Acquisition.} To capture fine-grained acoustic variations, we control three physical parameters during data collection: striker material (steel, wood, plastic), fill material (rice, water), and fill level (discretized into five states, where level 0 indicates a completely hollow object). We record synchronized impact audio (sampled at 48 kHz) and video (at $1024 \times 1024$ resolution) using a DJI Mic and a DJI Pocket 3 camera. To facilitate position-aware generation, the precise 3D spatial coordinates $(x, y, z)$ of each strike point are registered and serialized into structured JSON metadata, establishing a rigorous spatial-acoustic mapping for every sample.

\textbf{Acoustic Analysis.} To validate the physical consistency of our dataset, we analyze its fine-grained acoustic attributes (Figure \ref{fig:dataset_analy}). First, by rigorously isolating variables, specifically holding the container material, the striker, and the fill material constant to control for structural stiffness ($k$) and excitation conditions. We observe that the normalized dominant frequency strictly decreases as the internal mass ($m$) increases with higher fill levels. This aligns perfectly with the classical harmonic oscillator principle $f \propto \sqrt{k/m}$~\cite{fletcher1998physics}, providing a reliable prior for fill-level perception. Second, granular fillings (e.g., rice) exhibit significantly higher acoustic decay rates ($\alpha$) than liquids (e.g., water). This disparity is fundamentally driven by the high internal friction and collisional damping inherent in granular mechanics~\cite{jaeger1992physics}, ensuring strong acoustic discriminability for different internal states. Finally, harder impactors (e.g., steel) consistently excite higher-frequency resonance modes, yielding a significantly elevated spectral centroid compared to softer materials like wood or plastic. These distinctly preserved acoustic cues confirm that our dataset provides a rigorous foundation for accurate, physically grounded sound synthesis.

\section{Method}

\begin{figure*}
  \centering
  \includegraphics[width=0.95\textwidth]{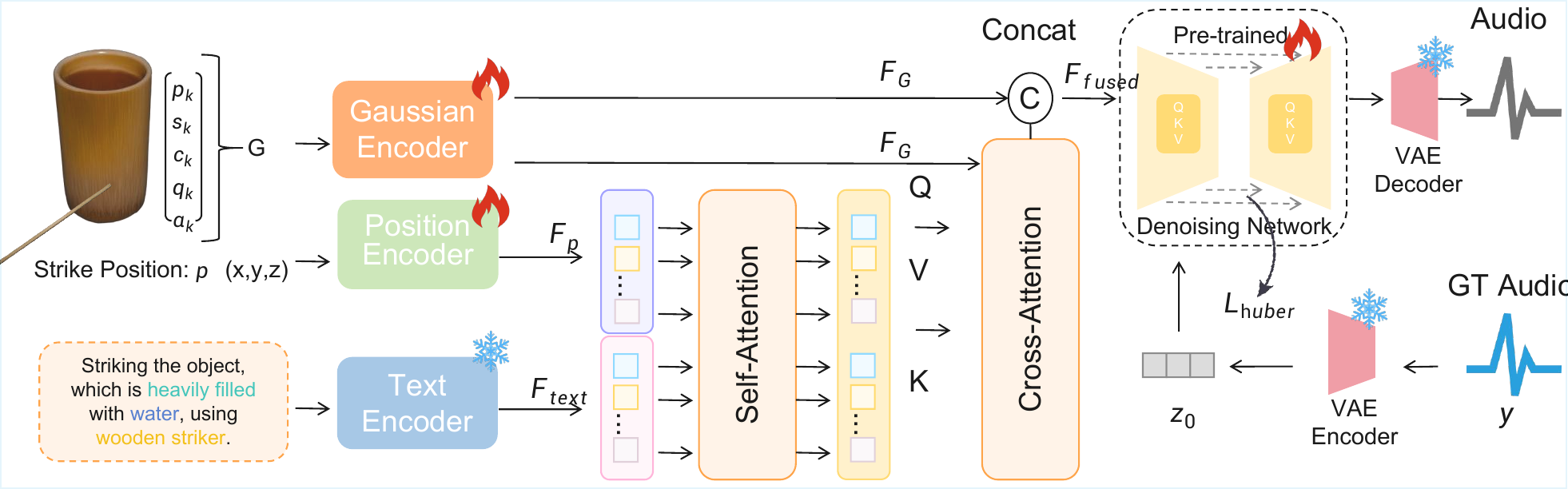}
  \caption{Overview of FillGauss. The framework first extracts individual features from the 3DGS representation $G$, the strike position $\mathbf{p}$, and the textual fill-state prompt. To accurately model physical interactions, a self-attention module combines the spatial and semantic inputs into a unified localized context. The global geometric features $\mathbf{F}_G$ then actively query ($\mathbf{Q}$) this context ($\mathbf{K}, \mathbf{V}$) through a cross-attention mechanism. The geometry-aware output is concatenated to form the definitive multi-modal condition $\mathbf{F}_{fused}$, which guides a Diffusion Transformer to denoise and recover the ground-truth audio latent $\mathbf{z}_0$. The entire latent diffusion process is optimized using a Huber loss $\mathcal{L}_{huber}$ to ensure high-fidelity impact sound synthesis.}
  \label{fig:pipeline_3}
\end{figure*}



\subsection{Problem Definition}

We represent the target object using 3DGS~\cite{kerbl20233dgs}, an explicit geometric formulation that is parameterized by a set of Gaussians
$G = \{\mathbf{p}_k, \mathbf{s}_k, \mathbf{c}_k, \mathbf{q}_k, \alpha_k\}_{k=1}^N,$
denoting the center position, scale, view-dependent color, rotation, and opacity, respectively. Motivated by our preceding dataset analysis, which proves that variations in internal states yield highly distinguishable acoustic signatures. We formalize the novel task of \textit{Fine-Grained Filling-Aware Impact Sound Generation}. Formally, given the 3DGS geometry $G$, the exact 3D strike position $\mathbf{p} \in \mathbb{R}^3$, the striker material, and the internal fill state (specifically, the fill material and continuous fill level), our objective is to synthesize a physically plausible impact audio $y$.

\textbf{Overview of FillGauss.}
To tackle this task, FillGauss jointly reasons over the visual, spatial, and internal physical states of an object. The framework takes three heterogeneous inputs: (1) the 3DGS representation $G$, (2) the 3D strike coordinates $\mathbf{p} \in \mathbb{R}^3$, and (3) a text prompt $\mathcal{T}$ explicitly detailing the striker and internal fill state. As shown in Figure~\ref{fig:pipeline_3}, a cross-modal attention fusion seamlessly entangles the geometric priors with the localized physical striking context. The resulting fused embedding then acts as the cross-attention condition for a Denoising Network (built upon TangoFlux), guiding the iterative latent diffusion process. Finally, a pre-trained Audio VAE decoder reconstructs the denoised latent back into the target audio waveform.

\subsection{Multi-modal Encoder}
To process the disparate input modalities, we employ dedicated encoders to extract robust semantic and geometric representations. 

\textbf{3DGS Encoder.} For the visual-geometric input, we adopt the 3DGS encoder architecture from SplatFormer~\cite{chen2024splatformer} to extract local point-wise features. The Gaussian parameters are projected into a sequence of features $\mathbf{F}_{G} \in \mathbb{R}^{N \times D}$, where $N$ is the number of Gaussian blobs and $D$ is the embedding dimension.

\textbf{Position Encoder.} The 3D impact coordinate $\mathbf{p} \in \mathbb{R}^3$ is crucial for determining the local resonance of the outer shell. We first apply a NeRF-style high-frequency positional encoding $\gamma(\mathbf{p})$:
\begin{equation}
\gamma(\mathbf{p}) = \left[ \sin(2^0 \pi \mathbf{p}), \cos(2^0 \pi \mathbf{p}), \dots, \sin(2^{L-1} \pi \mathbf{p}), \cos(2^{L-1} \pi \mathbf{p}) \right]
\end{equation}
The encoded vector is then mapped through a Multi-Layer Perceptron (MLP) to obtain the spatial feature $\mathbf{F}_{p} \in \mathbb{R}^{1 \times D}$.

\textbf{State Encoder.} To represent the fine-grained internal fill state (e.g., striker type, fill material, and fill level), we evaluated two distinctive encoding strategies (as compared in our ablation studies):

(1) \textit{HardCode Encoder:} We explicitly mapped the discrete categorical variables into learnable parameter spaces: 
\begin{equation}
    \mathbf{F}_{cond} = \text{Concat}(\text{Emb}_{s}(s), \text{Emb}_{m}(m), \text{Emb}_{l}(l)) \in \mathbb{R}^{3 \times D}
\end{equation}
where $s \in \mathcal{S}$, $m \in \mathcal{M}$, and $l \in \mathcal{L}$ denote the categorical indices for the striker type, fill material, and fill level, respectively.

(2) \textit{Fine-grained Text Prompting (Ours):} We translate the physical state into a language description $\mathcal{T}$ (e.g., "The object is hit with a steel striker. It is completely full of water.") and extract its sequence features $\mathbf{F}_{text} \in \mathbb{R}^{L_{text} \times D}$ using a pre-trained language model. 

As shown in Table~\ref{tab:ablation}, fine-grained text prompting yields superior acoustic fidelity. This naturally stems from our base model (TangoFlux) being a native text-to-audio architecture.

\subsection{Cross-modal Attention Fusion}

The core challenge in impact sound synthesis is accurately modeling the physical coupling between the impact location and the global acoustic system (container geometry and internal damping medium). To achieve this, we design a two-stage fusion module comprising Context Self-Attention and Geometry Cross-Attention.

First, to deeply entangle the internal semantic state with the precise spatial striking location, we concatenate the position embedding and the text embedding to form the context tokens $\mathbf{C} = [\mathbf{F}_{p} \parallel \mathbf{F}_{text}]$. We apply a Multi-Head Self-Attention (MHSA) layer to construct a unified physical context representation $\mathbf{C}'$:
\begin{equation}
\mathbf{C}' = \text{MHSA}(\mathbf{Q}=\mathbf{C}, \mathbf{K}=\mathbf{C}, \mathbf{V}=\mathbf{C}) + \mathbf{C}
\end{equation}
From a physical perspective, this self-attention mechanism acts as an implicit solver for the initial excitation state. It allows the model to globally reason about how the specific impact location interacts with the internal filling (e.g., striking the water line induces different initial energy compared to striking the hollow part above it).

Subsequently, to inject this localized physical context into the global 3D structural priors, we introduce a Geometry Cross-Attention module. Here, the dense 3D Gaussian features $\mathbf{F}_{G}$ serve as the queries ($\mathbf{Q}$), while the unified physical context $\mathbf{C}'$ acts as the keys ($\mathbf{K}$) and values ($\mathbf{V}$):
\begin{equation}
\mathbf{A}_{cross} = \text{MHCA}(\mathbf{Q}=\mathbf{F}_{G}, \mathbf{K}=\mathbf{C}', \mathbf{V}=\mathbf{C}')
\end{equation}
From a physical modeling perspective, this configuration allows the global 3D geometry to actively query the localized striking condition. Each Gaussian spatial region dynamically aggregates the impact properties (e.g., absorbing the semantic damping and positional excitation) based on its structural relevance to the strike. 

Finally, to ensure the denoising network preserves the raw, high-resolution 3D geometric topology while incorporating the conditionally activated acoustic context, we concatenate the original Gaussian features $\mathbf{F}_{G}$ with the cross-attention output $\mathbf{A}_{cross}$:
\begin{equation}
\mathbf{F}_{fused} = \text{Concat}(\mathbf{F}_{G}, \mathbf{A}_{cross})
\end{equation}
The resulting comprehensive multi-modal embedding, $\mathbf{F}_{fused}$, encapsulates the precise striking dynamics, the internal filling semantics, and the globally modulated 3D geometry. This fused representation is then fed into the subsequent network as the definitive cross-attention condition for the latent audio diffusion process.

\subsection{Latent Audio Diffusion and Optimization}

Instead of predicting high-dimensional audio waveforms directly, our framework is trained in the latent space of a pre-trained Audio VAE~\cite{evans2025stable} to balance computational efficiency and generation fidelity. Given a ground-truth audio waveform $y$, the VAE encoder compresses it into a highly structured latent representation $\mathbf{z}_{0} = \mathcal{E}_{audio}(y)$. During the forward diffusion process, Gaussian noise $\epsilon \sim \mathcal{N}(0, \mathbf{I})$ is progressively added to $\mathbf{z}_{0}$ based on a noise schedule $\sigma_{t}$ at timestep $t$, producing the noisy latent $\mathbf{z}_{t}$. 

Our Denoising Network $\mathcal{D}_{\theta}$ is built upon a Diffusion Transformer (DiT) architecture. The fused multi-modal embedding $\mathbf{F}_{fused}$ acts as the cross-attention condition inside the DiT blocks, effectively guiding the temporal denoising trajectory using the 3D geometric and physical priors. The network takes $\mathbf{z}_{t}$, $\mathbf{F}_{fused}$, and the timestep $t$ to predict the target velocity (or noise residual) $\mathbf{v} = \epsilon - \mathbf{z}_{0}$. 

To improve the robustness of the generation against outliers and stabilize the latent diffusion training, we employ the Huber Loss ($\mathcal{L}_{huber}$) instead of the standard Mean Squared Error (MSE):
\begin{equation}
\mathcal{L}_{diffusion} = \mathbb{E}_{\mathbf{z}_{0}, \epsilon, t} \left[ \frac{1}{B} \sum_{i=1}^{B} \mathcal{L}_{huber}(\mathcal{D}_{\theta}(\mathbf{z}_{t}, \mathbf{F}_{fused}, t), \mathbf{v}; \delta) \right]
\end{equation}
where $B$ is the batch size and $\delta$ is the threshold parameter (set to 1.0 in our experiments). The Huber loss elegantly acts as an $L_2$ loss for small errors (enhancing fine-grained spectrogram details) and behaves like an $L_1$ loss for large errors (preventing gradient explosions during the early training stages). During inference, the predicted clean latent $\mathbf{z}_{0}$ is passed through the VAE decoder to reconstruct the final impact sound.

\section{Experiments} 

\subsection{Implementation}

\textbf{Compared Methods.} To evaluate FillGauss, we compare it against two primary baselines: \textbf{SonicGauss}~\cite{wang2025sonicgauss} and \textbf{TangoFlux}~\cite{hung2024tangoflux}. SonicGauss is a state-of-the-art 3D-aware baseline that conditions a diffusion model on 3DGS but relies on generic captions, thereby failing to accurately capture complex internal filling states. TangoFlux is a highly capable pure text-to-audio generative architecture; while it excels at semantic audio generation, it fundamentally lacks 3D spatial grounding and cannot perform position-aware synthesis from visual geometry.

\textbf{Evaluation Details.}
To comprehensively evaluate the generated impact sounds, we employ both objective metrics and subjective human perceptual evaluations. 
\textbf{(1) Objective Metrics.} Our primary metric is the Fréchet Audio Distance (FAD)~\cite{kilgour2019frechet}, a reference-free evaluation metric that measures the distribution similarity between the generated audio features and the ground truth using a pre-trained VGGish model~\cite{hershey2017cnn}. A lower FAD indicates that the generated sounds are statistically closer to the real-world acoustic distribution. In addition, we report the Kullback-Leibler (KL) divergence~\cite{kullback1951information} to assess fine-grained semantic feature alignment between the generated spectrograms and the targets. Finally, we compute the Inception Score (IS)~\cite{salimans2016improved} to evaluate the overall clarity and diversity of the generated acoustic samples.
\textbf{(2) Subjective Metrics.} We conducted a blind perceptual study with 8 participants, evaluating four metrics. The \textbf{Mean Opinion Score (MOS)} measures overall audio naturalness on a 1–5 scale~\cite{itu800}. The \textbf{Win Rate (\%)}~\cite{ribeiro2011crowdsourcing} indicates the preference percentage in an A/B test against the baseline. To assess physical grounding, participants blindly identified five specific attributes (striker, container material, strike position, fill material, and fill level) per audio sample. From this, we report the \textbf{Matched Attr. Count} as the average number of correctly identified attributes (out of 5), and the \textbf{Fine-grained Attribute Matching Rate (\%)} as the individual identification accuracy for each specific property.

\textbf{Implementation Details.} To train and evaluate FillGauss, we partition FillImpact Dataset into an 80\% training split and a 20\% testing split. Exhaustive details regarding audio preprocessing (e.g., latent compression via Audio VAE~\cite{evans2025stable}) and the strict global bounding-box normalization of the 3DGS spatial coordinates are deferred to the Appendix. All models in our framework are implemented using PyTorch. We conduct our experiments on a single NVIDIA RTX 4090 GPU (24GB). During the training phase, we set the per-device batch size to 15 and train the model for 80 epochs utilizing the AdamW optimizer~\cite{loshchilov2019decoupled} with a base learning rate of $1 \times 10^{-4}$ and a linear learning rate scheduler. Building upon the foundational architecture of SonicGauss~\cite{wang2025sonicgauss}, our core denoising network is initialized by fine-tuning the pre-trained TangoFlux model~\cite{hung2024tangoflux}. To preserve their robust pre-trained priors, we maintain frozen weights for both the Audio VAE and the base Text Encoder~\cite{chung2022scaling,raffel2020exploring} throughout the training process.

\subsection{Results}

To comprehensively validate our framework against the novel task of fine-grained filling-aware sound generation, we structure our evaluation across three dimensions. We visually verify physical interactions (Qualitative), statistically measure generation fidelity (Quantitative), and assess human perceptibility (Subjective). Together, these experiments rigorously test our model's capacity for position-aware, striker-aware, and fill-aware synthesis.

\begin{figure}[htbp]
  \centering
  \includegraphics[width=0.45\textwidth]{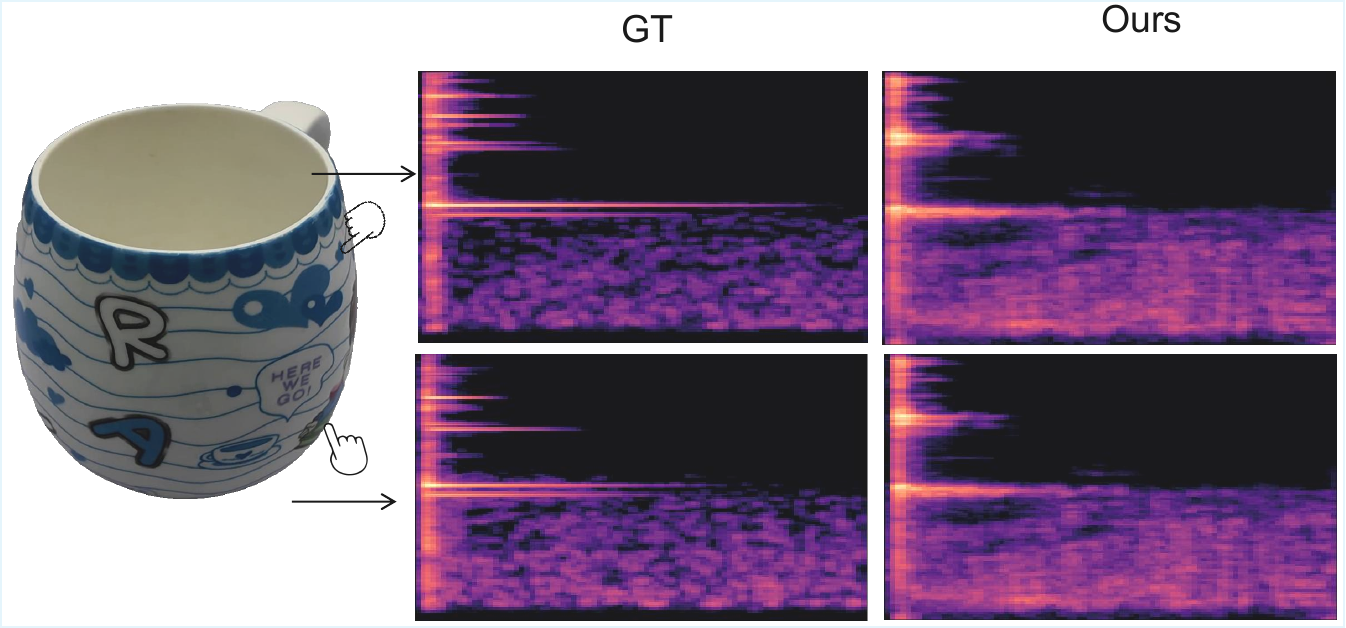}
  \caption{Results of position-aware sound generation.}
  \label{fig:experiment_position}
\end{figure}

\begin{figure}[htbp]
  \centering
  \includegraphics[width=0.4\textwidth]{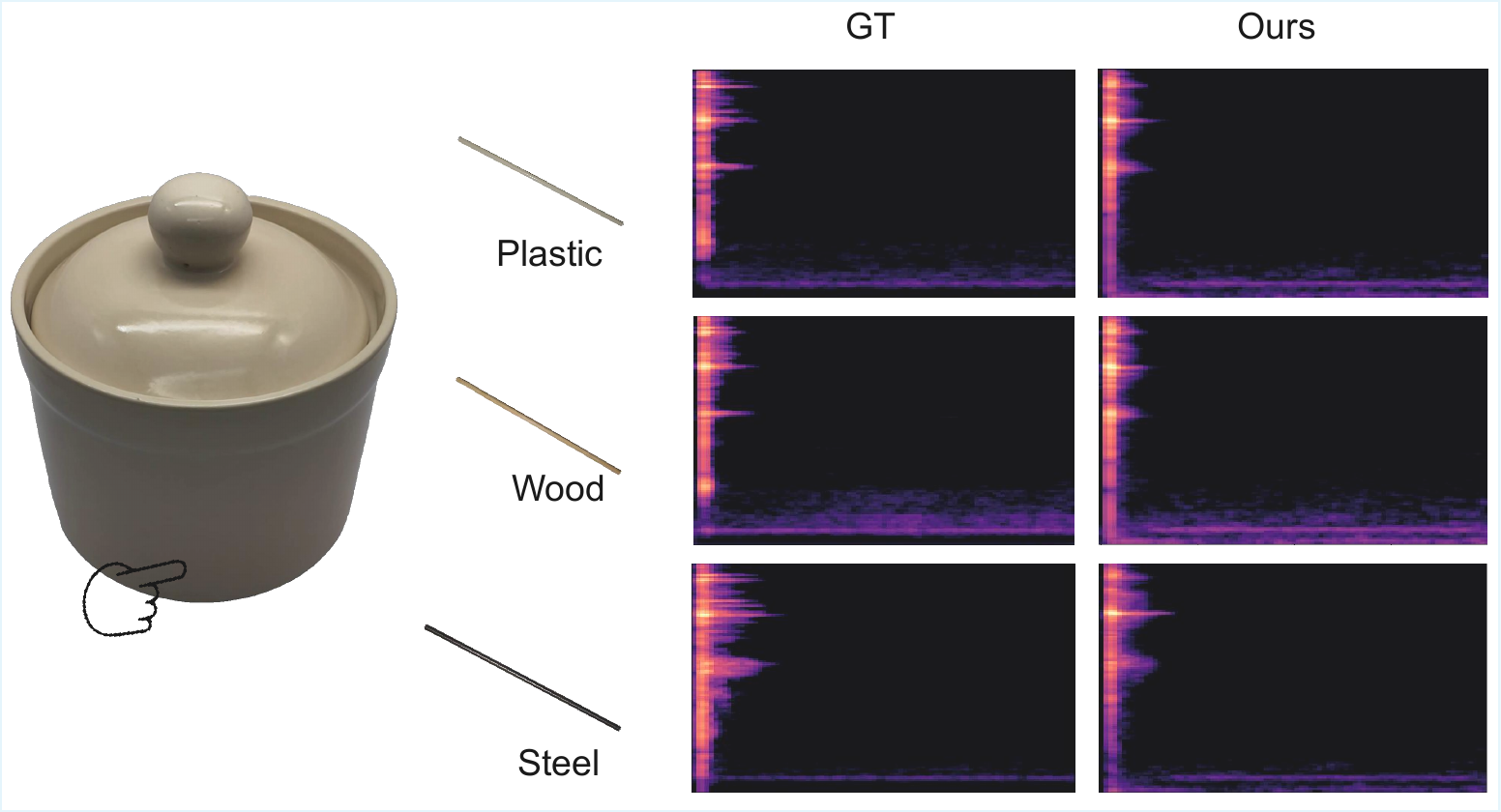}
  \caption{Results of striker-aware sound generation across different tool materials (plastic, wood, and steel).}
  \label{fig:experiment_striker}
\end{figure}

\begin{figure}[htbp]
  \centering
  \includegraphics[width=0.5\textwidth]{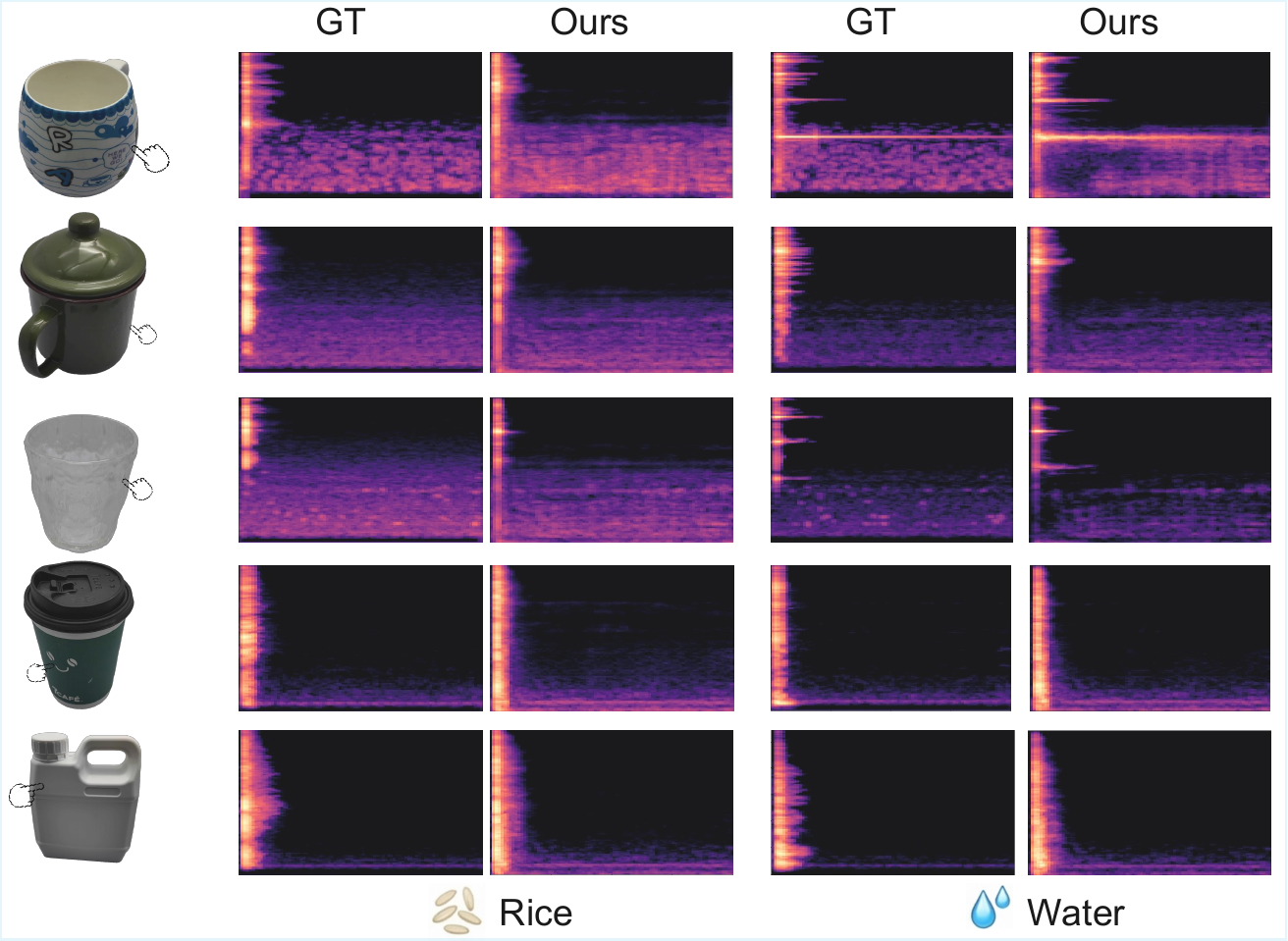}
  \caption{Results of fill-material-aware sound generation.}
  \label{fig:experiment_fillmat}
\end{figure}

\begin{figure*}[htbp]
  \centering
  \includegraphics[width=0.8\textwidth]{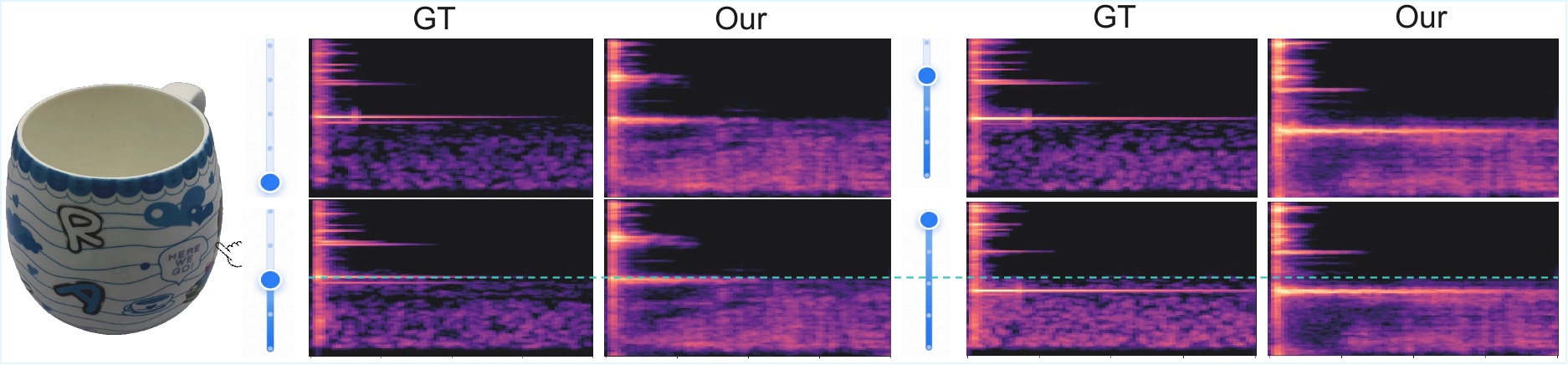}
  \caption{Qualitative results of fill-level-aware sound generation across continuous filling states.}
  \label{fig:experiment_level}
\end{figure*}

\textbf{Qualitative Results.} We analyze the generated spectrograms to verify FillGauss's ability to capture localized and internal physical variations. For position-awareness (Figure \ref{fig:experiment_position}), shifting the strike location across the 3DGS container reflects precise shifts in resonance modes and energy decay, demonstrating accurate spatial geometry learning. Regarding striker-awareness (Figure \ref{fig:experiment_striker}), the model accurately synthesizes distinct spectral energy distributions, correctly reflecting how harder materials like steel excite higher frequency bands than wood or plastic. Crucially, we validate the internal state reasoning central to our task. For fill-level awareness (Figure \ref{fig:experiment_level}), increasing the internal capacity visibly shifts the dominant frequency bands downwards. The blue dashed lines explicitly trace the physical rule established in our dataset analysis, firmly adhering to the added-mass harmonic oscillator principles ($f \propto \sqrt{k/m}$). Finally, evaluating fill-material awareness (Figure \ref{fig:experiment_fillmat}) reveals distinct acoustic signatures. FillGauss successfully synthesizes the rapid energy decay characteristic of granular damping for rice alongside the prolonged resonance typical of liquid water.

\textbf{Quantitative Results.} Table \ref{tab:quantitative} compares FillGauss against the 3D-aware SonicGauss and the text-to-audio TangoFlux baselines. Among 3D-aware frameworks, FillGauss achieves the best primary metric performance (FAD 1.3500), significantly outperforming the fine-tuned SonicGauss. The poor zero-shot performance of SonicGauss exposes a critical flaw in existing methods. Without explicit internal filling knowledge, models average the acoustic outputs of identical-looking containers with vastly different acoustic responses, causing severe distribution mismatch. While the fine-tuned TangoFlux achieves a lower FAD due to its unconstrained text-to-audio architecture, it fundamentally lacks 3D geometric perception and cannot associate sounds with precise spatial coordinates. FillGauss achieves the optimal balance, establishing the state-of-the-art for physically grounded and spatially-aware audio generation.

\begin{table}[htbp]
\centering
\setlength{\tabcolsep}{3.5pt} 
\caption{Quantitative comparison of impact sound generation. ``ZS'' and ``FT'' denote Zero-shot and Fine-tuned settings. $\downarrow$ indicates lower is better, $\uparrow$ indicates higher is better. \textbf{Bold} denotes best, \underline{underline} denotes second best.}
\label{tab:quantitative}
\begin{tabular}{lcccc}
\toprule
\textbf{Model} & \textbf{FAD} $\downarrow$ & \textbf{KL sig} $\downarrow$ & \textbf{IS Avg} $\uparrow$ & \textbf{IS Std} $\downarrow$ \\
\midrule
SonicGauss (ZS) & 7.8909 & 1.2022 & 1.0136 & 1.85$\times10^{-4}$ \\
TangoFlux (ZS) & 15.5368 & 1.6947 & 1.0062 & \textbf{0.85$\times10^{-4}$} \\
\midrule
SonicGauss (FT) & 1.5653 & \textbf{0.5750} & \underline{1.0156} & 2.85$\times10^{-4}$ \\
TangoFlux (FT) & \textbf{1.0627} & 0.7700 & 1.0151 & \underline{2.61$\times10^{-4}$} \\
\midrule
\textbf{FillGauss (Ours)} & \underline{1.3500} & \underline{0.6977} & \textbf{1.0160} & 2.93$\times10^{-4}$ \\
\bottomrule
\end{tabular}
\end{table}


\textbf{Human Perceptual Evaluation.} As shown in Table \ref{tab:user_study_main}, FillGauss securing a decisive majority in the overall preference win rate. More importantly, Table \ref{tab:user_study_fine} details the Fine-grained Attribute Matching Rate. Even when fine-tuned, the baselines struggle significantly to convey internal states, notably poor perception accuracy for both fill material and fill level. In contrast, FillGauss successfully transmits complex physical conditions and achieves the highest perception accuracy across all evaluated categories including striker material, container material, strike position, fill material, and fill level. Although a gap remains toward high-fidelity human perception, our approach significantly outperforms existing baseline methods.

\begin{table}[htbp]
\centering
\setlength{\tabcolsep}{3.5pt} 
\caption{Comprehensive Subjective Evaluation Results. ``ZS'' and ``FT'' denote Zero-shot and Fine-tuned. $\uparrow$ indicates higher is better. \textbf{Bold} denotes the best performance.}
\label{tab:user_study_main}
\begin{tabular}{lccc}
\toprule
\textbf{Model} & \textbf{MOS} ($\uparrow$) & \textbf{Win Rate} ($\uparrow$) & \textbf{Attr. Count} ($\uparrow$) \\
\midrule
TangoFlux (ZS) & 1.50 $\pm$ 1.16 & 3.6\% & 0.36 $\pm$ 0.98 \\
TangoFlux (FT) & 3.64 $\pm$ 0.74 & 37.5\% & 2.64 $\pm$ 1.48 \\
SonicGauss (ZS) & \textbf{4.50 $\pm$ 0.65} & 16.1\% & 1.43 $\pm$ 1.42 \\
SonicGauss (FT) & 4.36 $\pm$ 0.63 & 33.9\% & 1.95 $\pm$ 1.77 \\
\midrule
\textbf{FillGauss (Ours)} & 4.43 $\pm$ 0.76 & \textbf{69.6\%} & \textbf{3.62 $\pm$ 1.51} \\
\bottomrule
\end{tabular}
\end{table}

\begin{table}[htbp]
\centering
\setlength{\tabcolsep}{2.5pt} 
\caption{Fine-grained Attribute Matching Rate (\%). ``ZS'' and ``FT'' denote Zero-shot and Fine-tuned settings. \textbf{Bold} denotes the best performance.}
\label{tab:user_study_fine}
\begin{tabular}{lccccc}
\toprule
\textbf{Model} & \textbf{Striker} & \textbf{Mat.} & \textbf{Pos.} & \textbf{Fill Mat.} & \textbf{Level} \\
\midrule
TangoFlux (ZS) & 11.9\% & 7.1\% & 9.5\% & 4.8\% & 2.4\% \\
TangoFlux (FT) & 73.8\% & 71.4\% & 59.5\% & 31.0\% & 28.6\% \\
SonicGauss (ZS) & 50.0\% & 21.4\% & 31.0\% & 14.3\% & 26.2\% \\
SonicGauss (FT) & 54.8\% & 35.7\% & 42.9\% & 31.0\% & 31.0\% \\
\midrule
\textbf{FillGauss (Ours)} & \textbf{88.1\%} & \textbf{90.5\%} & \textbf{71.4\%} & \textbf{52.4\%} & \textbf{59.5\%} \\
\bottomrule
\end{tabular}
\end{table}

\subsection{Ablation Study}

To validate our network components, we conducted an ablation study (Table \ref{tab:ablation}), prioritizing the Fréchet Audio Distance (FAD) as the primary metric for overall acoustic realism. Our full architecture (\textbf{Final}) achieves the best FAD, whereas removing the cross-attention module (\textit{w/o cross-attention}) causes the most severe degradation. Without it, the model becomes blind to the 3DGS geometry and overfits to explicit text conditions; this pathological shortcut learning generates generic template sounds that artificially inflate secondary metrics (KL and IS) but catastrophically destroy physical realism. Similarly, omitting the self-attention module (\textit{w/o self-attention}) prevents the smooth entanglement of spatial impact positions with internal filling semantics, failing to reason globally about physical interactions. Furthermore, replacing the Huber loss with a standard MSE loss (\textit{w/o $\mathbf{L}_{huber}$}) reduces training robustness against early-stage diffusion noise, worsening the FAD. Finally, substituting fine-grained text prompts with discrete categorical embeddings (\textit{Hard-code}) noticeably drops performance, confirming that natural language optimally leverages the rich semantic priors of the pre-trained text encoder to achieve physically grounded sound generation. 
\begin{table}[htbp]
\centering
\caption{Ablation study on the core components of FillGauss. $\downarrow$ indicates lower is better, $\uparrow$ indicates higher is better. \textbf{Bold} denotes the best performance.}
\label{tab:ablation}
\begin{tabular}{lcccc}
\toprule
\textbf{Configuration} & \textbf{FAD} $\downarrow$ & \textbf{KL sig} $\downarrow$ & \textbf{IS Avg} $\uparrow$ & \textbf{IS Std} $\downarrow$ \\
\midrule
w/o self-attention & 1.4111 & 0.6434 & 1.0161 & 2.89$\times10^{-4}$ \\
w/o cross-attention & 1.5623 & \textbf{0.5904} & \textbf{1.0162} & 2.84$\times10^{-4}$ \\
w/o $\mathbf{L}_{huber}$ & 1.4854 & 0.6297 & \textbf{1.0162} & 3.01$\times10^{-4}$ \\
Hard-code & 1.4936 & 0.6506 & 1.0158 & \textbf{2.81$\times10^{-4}$} \\
\midrule
\textbf{Final (Ours)} & \textbf{1.3500} & 0.6977 & 1.0160 & 2.93$\times10^{-4}$ \\
\bottomrule
\end{tabular}
\end{table}



\subsection{Sensitivity Analysis}

\begin{figure}[htbp]
  \centering
  \includegraphics[width=1\linewidth]{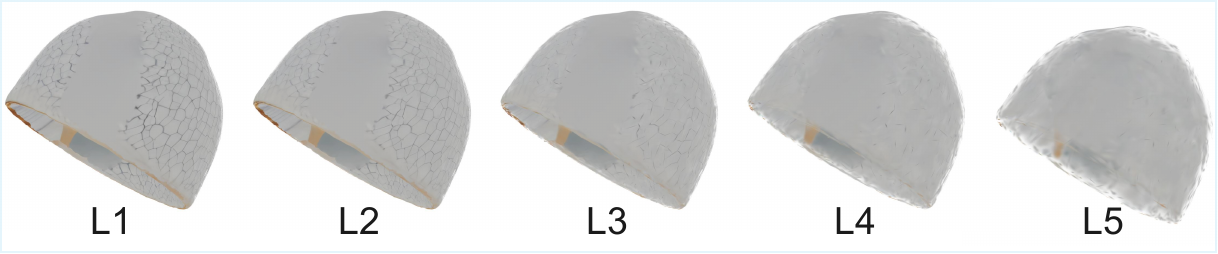} 
  \caption{Visual comparison of 3DGS reconstructions under varying sparsity levels. Further sparsification (L3 to L5) introduces severe artifacts like semi-transparent floaters, degrading acoustic performance.}
  \label{fig:sparsity_visuals}
\end{figure}

To verify that FillGauss genuinely relies on the fine-grained material priors of the 3DGS input rather than overfitting to the text prompt, we analyze its sensitivity to varying sparsity levels (Table \ref{tab:ablation_sparsity}). To systematically construct these levels (L1 to L5), we apply a physical energy-based voxel downsampling strategy with increasing voxel sizes $v \in \{0.0, 0.02, 0.05, 0.08, 0.12\}$. Within each populated voxel $V$, we retain only the single Gaussian $k^*$ that maximizes an energy score $E_k = \sum_{i \in \{x,y,z\}} \log(s_{k,i}) + \alpha_k$. This formulation explicitly prioritizes Gaussians with significant volumetric influence and high opacity while strictly preserving their original geometric parameters. As shown in Figure \ref{fig:sparsity_visuals}, the reconstructions maintain solid surface integrity and clear textures from L1 to L3, corresponding to a graceful and stable decline in acoustic fidelity. This demonstrates our framework's robustness to moderate compression. However, under extreme sparsification (L4 and L5), while the macro geometric shape remains broadly discernible, the surface integrity and texture degrade significantly. The object boundary transitions into a semi-transparent cloud of floaters, losing its explicit material color ($\mathbf{c}_k$) and structural density ($\alpha_k$). Crucially, this specific loss of high-frequency surface details explicitly triggers a significant drop in audio performance (FAD worsens to 2.5718). Consequently, L3 emerges as an optimal boundary balancing computational efficiency and acoustic realism, demonstrating that our framework can successfully utilize a significantly reduced number of Gaussians to synthesize relatively high-quality impact sounds.

\begin{table}[htbp]
\centering
\setlength{\tabcolsep}{3.5pt} 
\caption{Impact of 3DGS sparsity levels (L1 to L5) on sound generation. $\downarrow$ indicates lower is better, $\uparrow$ indicates higher is better. \textbf{Bold} denotes the best performance.}
\label{tab:ablation_sparsity}
\begin{tabular}{lcccc}
\toprule
\textbf{Configuration} & \textbf{FAD} $\downarrow$ & \textbf{KL sig} $\downarrow$ & \textbf{IS Avg} $\uparrow$ & \textbf{IS Std} $\downarrow$ \\
\midrule
\textbf{L1 (Ours)} & \textbf{1.3504} & 0.6977 & 1.0160 & 2.93$\times10^{-4}$ \\
L2 & 1.3940 & 0.6872 & 1.0161 & 2.85$\times10^{-4}$ \\
L3 & 1.6182 & 0.6712 & 1.0166 & \textbf{2.68$\times10^{-4}$} \\
L4 & 2.2265 & \textbf{0.6642} & 1.0172 & 2.88$\times10^{-4}$ \\
L5 & 2.5718 & 0.7439 & \textbf{1.0175} & 2.81$\times10^{-4}$ \\
\bottomrule
\end{tabular}
\end{table}

\section{Conclusion}

We introduce FillGauss and FillImpact dataset to generate fine-grained, filling-aware impact sounds for 3DGS representations. By fusing 3DGS, strike position, and fill conditions, the framework generates acoustics that adhere to established physical principles.
Despite these contributions, certain limitations remain. \textbf{Limitations:} Extreme 3DGS sparsification degrades surface integrity and compromises acoustic fidelity, and human perceptual evaluations indicate a remaining gap toward high-fidelity auditory distinguishability. In the future, we will focus on extending the model to capture complex internal dynamics, such as sloshing fluids and soft-body deformations.

\bibliographystyle{IEEEtran}
\bibliography{main}

\clearpage
\appendix
%

\section{Experiments}

\subsection{Data Processing Details}

All ground truth audio recordings are resampled to 48 kHz and dynamically padded or truncated to a fixed maximum duration to match the input sequence requirements of the latent diffusion model. The raw audio is then compressed into a lower-dimensional latent space using the pre-trained Audio VAE from Stable Audio Open~\cite{evans2025stable}. 

For the visual modality, real-world 3DGS reconstructions contain arbitrary physical scales, whereas our 3DGS encoder requires strictly normalized inputs for stable feature extraction. To ensure spatial consistency and scale-invariance, we perform a global bounding-box normalization on the 3DGS representations and their corresponding strike coordinates. Given the set of 3D Gaussian positions $\mathcal{X} = \{\mathbf{x}_i\}_{i=1}^N$, we compute the bounding box extremes $\mathbf{x}_{min}$ and $\mathbf{x}_{max}$. The geometric center $\mathbf{c}$ and the maximum spatial extent $E_{max}$ are defined as:
\begin{equation}
\mathbf{c} = \frac{\mathbf{x}_{max} + \mathbf{x}_{min}}{2}, \quad E_{max} = \max(\mathbf{x}_{max} - \mathbf{x}_{min})
\end{equation}
To strictly confine the object within a coordinate range of $[-1.5, 1.5]^3$, we calculate a global scale factor $S = 3 / E_{max}$. Each Gaussian's position $\mathbf{x}_i$ and the precise 3D strike coordinate $\mathbf{p}$ are synchronously translated to the origin and scaled:
\begin{equation}
\mathbf{x}'_i = S(\mathbf{x}_i - \mathbf{c}), \quad \mathbf{p}' = S(\mathbf{p} - \mathbf{c})
\end{equation}
Furthermore, because the 3D Gaussian scale parameters $\mathbf{s}_i \in \mathbb{R}^3$ are optimized and stored in logarithmic space, we apply a necessary additive shift to synchronize their sizes with the new coordinate space:
\begin{equation}
\mathbf{s}'_{i} = \mathbf{s}_{i} + \ln(S)
\end{equation}
This rigorous normalization guarantees that all geometric priors and interaction points share a unified, normalized spatial distribution, significantly stabilizing the cross-modal attention mechanisms during training.

\subsection{Baseline Fine-Tuning Details}

To ensure a fair and rigorous comparison, we fine-tune the selected baselines on our FillImpact dataset, meticulously adapting the data formats to match their distinct architectural requirements.

\subsubsection{SonicGauss Fine-Tuning}
The SonicGauss framework is natively designed to synthesize sound conditioned on 3D Gaussian Splatting representations and spatial impact coordinates. Consequently, we directly utilize the 3DGS models, precise strike positions, and the corresponding ground-truth audio recordings from our dataset to fine-tune this baseline, maintaining its original modality alignment without requiring additional data transformations.

\subsubsection{TangoFlux Fine-Tuning}
Unlike models utilizing explicit 3D visual conditions, TangoFlux operates strictly as a text-to-audio generation framework. To leverage our entire dataset for fine-tuning TangoFlux, we systematically transform all multimodal physical parameters into rich textual prompts. The prompt construction begins with a base visual caption describing the target object. We then semantically translate the exact strike coordinates into descriptive text, mapping specific points to relative spatial locations such as the lower part, the middle part, or the upper part. 

Furthermore, the complex internal physical states are explicitly converted into natural language descriptions. Objects without fill are described as hollow empty containers or completely solid objects depending on their baseline state. For filled objects, we apply proportional text templates to articulate the exact fill level, generating phrases ranging from quarter full to completely full of a specific material. The final training prompt seamlessly concatenates the visual caption, the designated strike location, the striker material, and the internal state description. This comprehensive text condition ensures that TangoFlux receives the equivalent physical context provided to our multimodal framework during the fine-tuning process.

\subsection{Evaluation Details}
\textbf{Subjective Metrics.} For subjective assessment, we conducted a blind human perceptual evaluation with 20 participants using high-quality headphones. We utilize four specific metrics to evaluate the generated audio across different perceptual dimensions:

\begin{itemize}

    \item \textbf{Mean Opinion Score (MOS):} To evaluate overall audio naturalness and physical plausibility, participants rated the generated audio on a standard scale from 1 (completely unnatural) to 5 (highly realistic).

    \item \textbf{Win Rate (\%):} We conducted an A/B preference test, presenting paired audio samples (Our model vs. Baseline) side-by-side. The Win Rate represents the percentage of trials where our generated audio was strictly preferred by the listener.

    \item \textbf{Matched Attr. Count:} We designed a blind Fine-grained Attribute Matching test where participants listened to the audio without any visual cues and attempted to identify five specific physical properties: the striker material, container material, strike position, fill material, and fill level. This metric reports the average number of correctly identified attributes per sample (out of a maximum of 5), presented as mean $\pm$ standard deviation.

    \item \textbf{Fine-grained Attribute Matching Rate (\%):} Derived from the same blind test, this metric breaks down the perceptual accuracy for each individual physical property. It calculates the exact percentage of instances where participants correctly perceived a specific attribute (e.g., distinguishing water from rice, or wood from steel) solely from the acoustic signals.

\end{itemize}

\section{FillImpact}

\begin{figure*}[htbp]
  \centering
  \includegraphics[width=\textwidth]{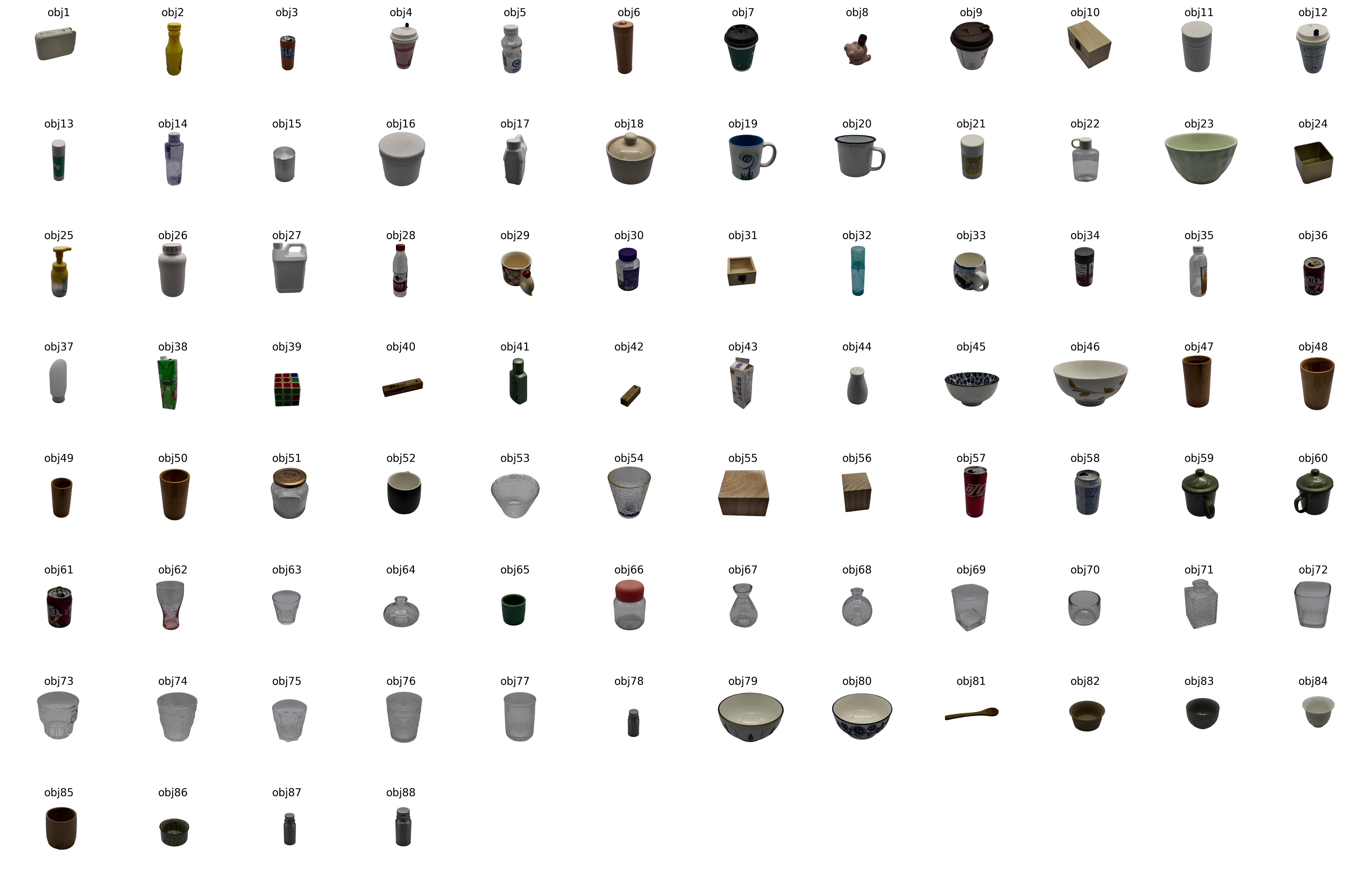}
  \caption{An overview of the 88 real-world objects in FillImpact, meticulously curated to capture the coupled acoustic dynamics of 3D geometry and varying internal filling states.}
  \label{fig:dataset_objects_overview}
\end{figure*}

To fundamentally address the limitations of existing datasets that strictly assume hollow or solid objects, we construct FillImpact. It is a pioneering multi-modal collection specifically designed to capture the complex acoustic dynamics of internal fillings coupled with 3D object geometry. As illustrated in Figure \ref{fig:dataset_objects_overview}, the dataset comprises 88 diverse, real-world everyday containers, yielding a total of over 5,000 rigorous acoustic recordings.

\begin{figure}[htbp]
  \centering
  \includegraphics[width=0.4\textwidth]{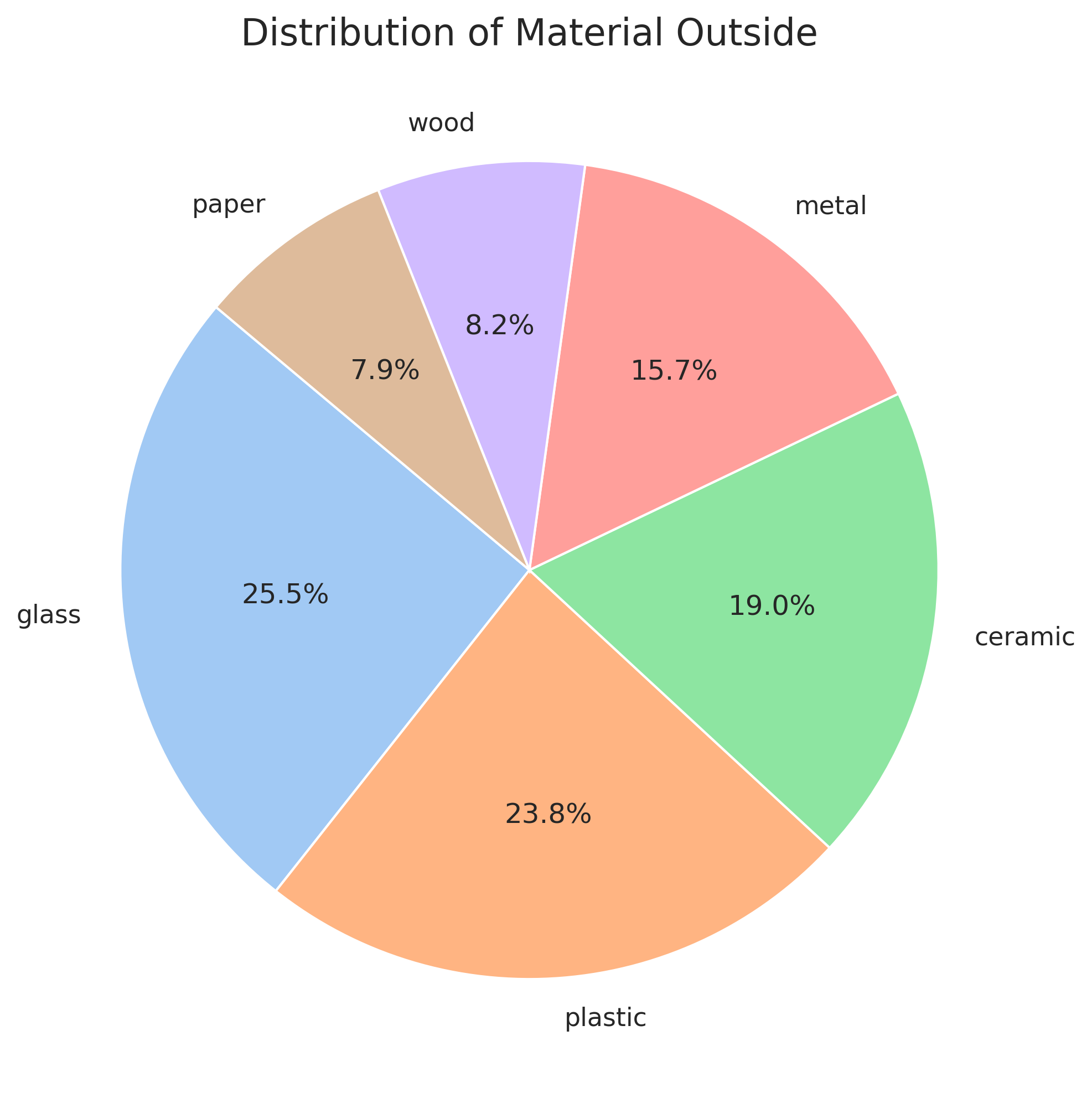}
  \caption{Distribution of audio recordings across the six external container material categories, ensuring a diverse range of structural resonance priors.}
  \label{fig:01_material_outside_pie}
\end{figure}

\begin{figure}[htbp]
  \centering
  \includegraphics[width=0.4\textwidth]{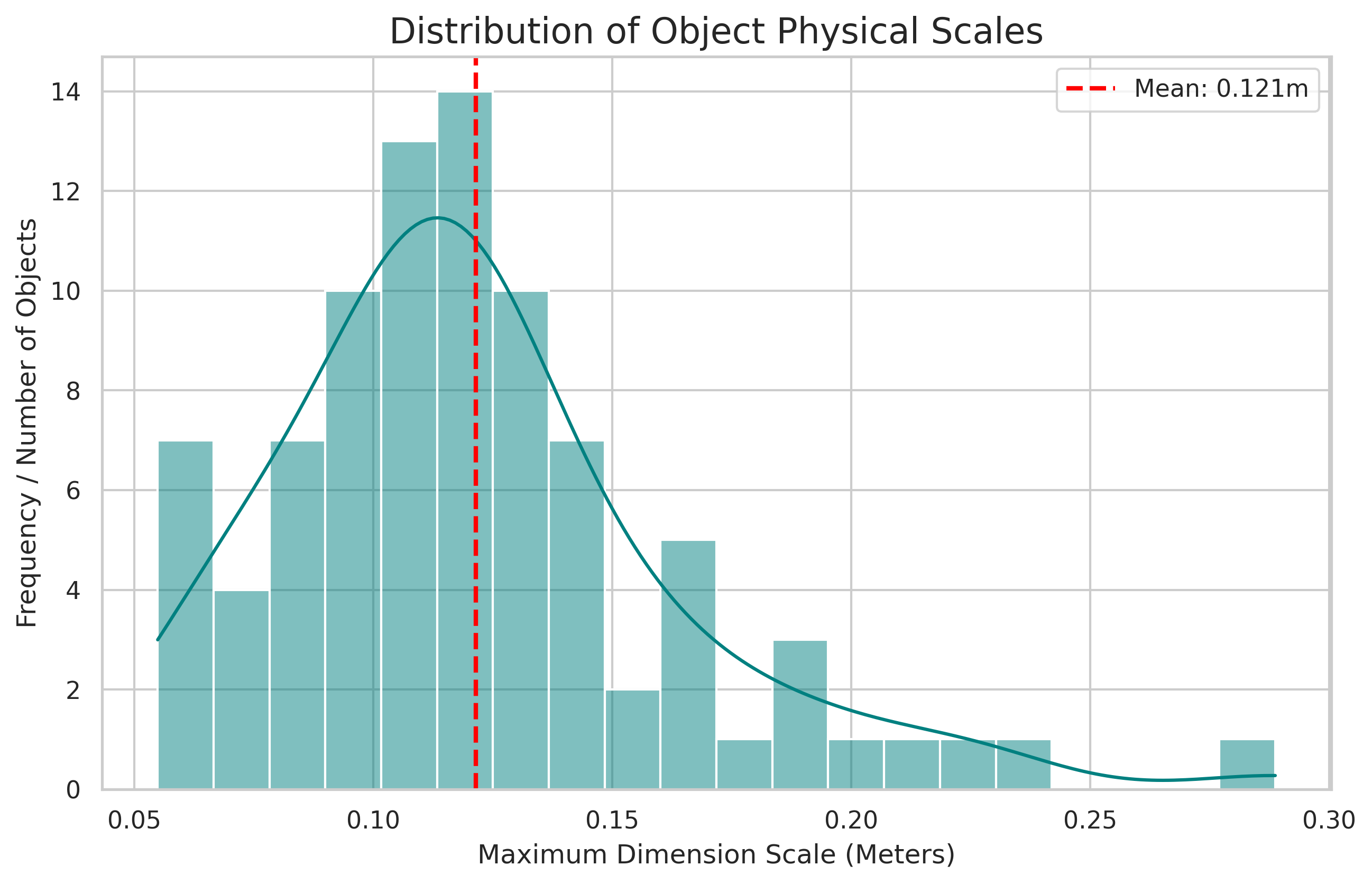}
  \caption{Distribution of physical scales across all collected objects, demonstrating the geometric diversity of the resonant cavities.}
  \label{fig:06_object_scale_distribution}
\end{figure}

\textbf{Object and Material Diversity.} 
Ensuring broad physical and geometric diversity is crucial for the generalization of our generative model. Figure \ref{fig:01_material_outside_pie} details the distribution of the collected audio recordings across the external container materials, covering six distinct common categories (glass, plastic, ceramic, metal, wood, and paper). This sample diversity guarantees that our model learns a wide spectrum of baseline structural resonances. Furthermore, to accurately capture acoustic variations related to the physical size of the resonant cavity, Figure \ref{fig:06_object_scale_distribution} visualizes the statistical distribution of the physical scales across all 88 objects. The maximum dimension scales exhibit a normal-like distribution (mean $\approx$ 0.12m), reflecting realistic sizes from small cups to larger storage bins.

\begin{figure}[htbp]
  \centering
  \includegraphics[width=0.4\textwidth]{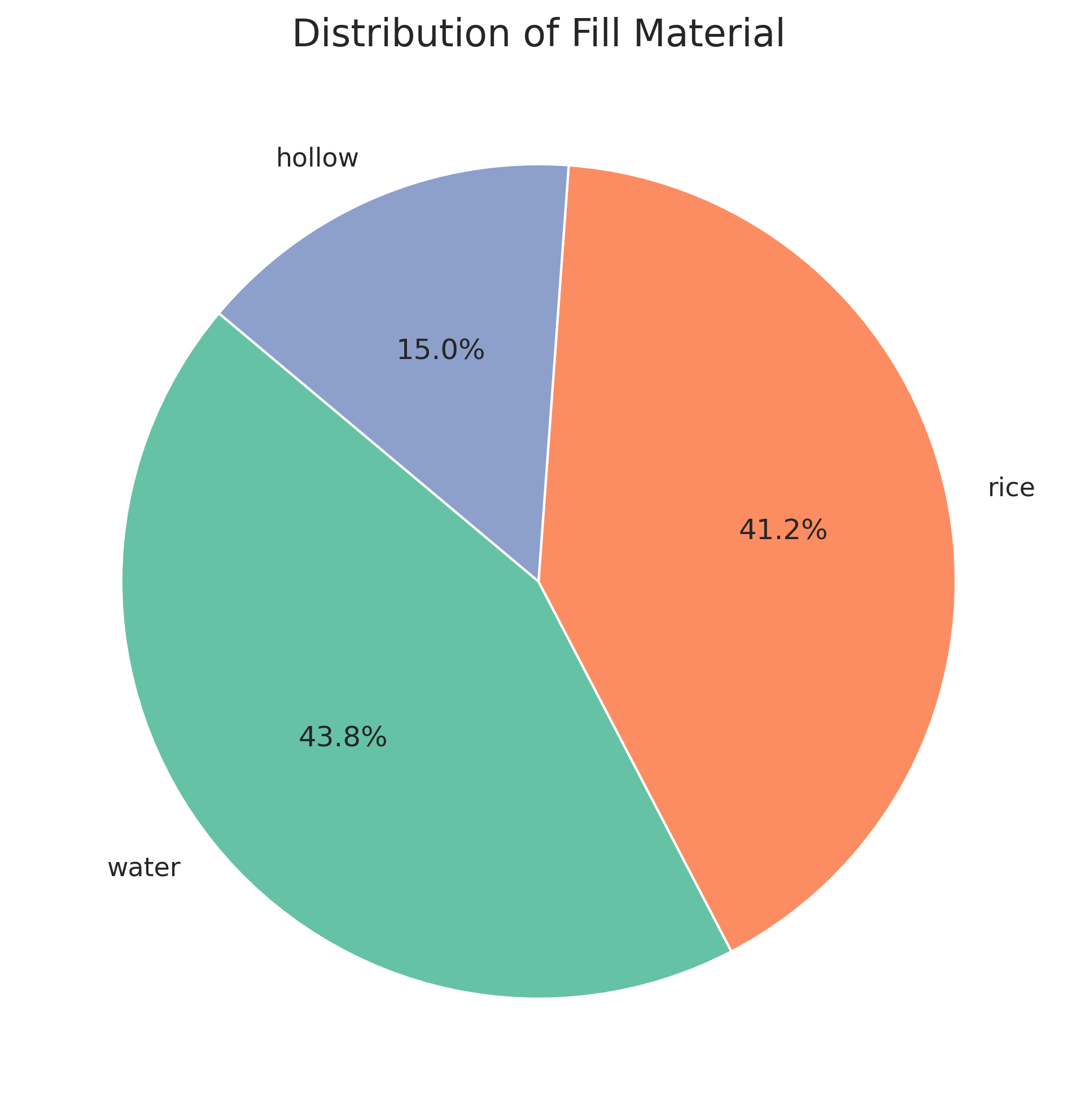}
  \caption{Proportional distribution of audio recordings based on the internal fill states (i.e., granular rice, liquid water, and hollow baselines).}
  \label{fig:02_fill_material_pie_updated}
\end{figure}

\begin{figure}[htbp]
  \centering
  \includegraphics[width=0.4\textwidth]{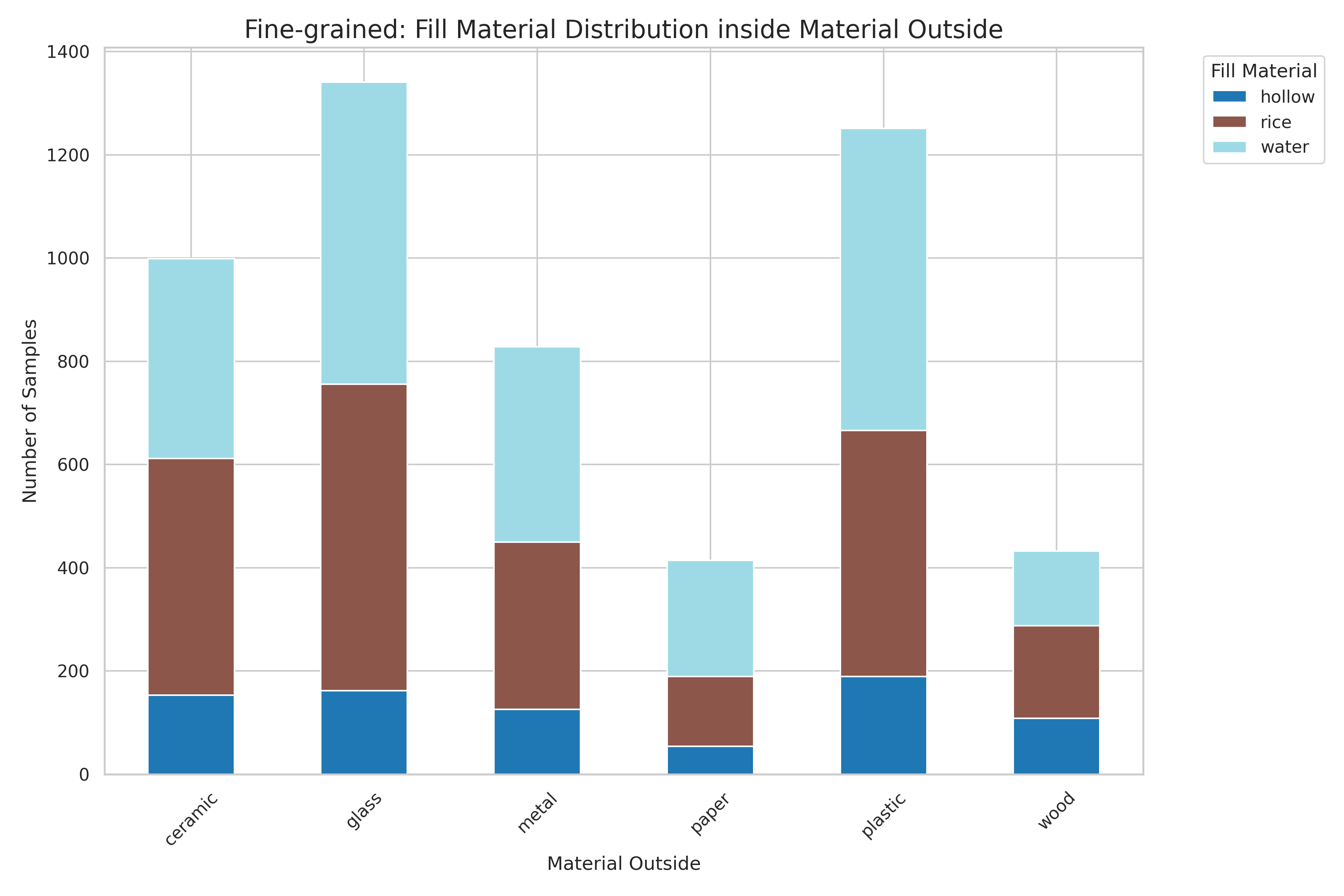}
  \caption{Stacked bar chart illustrating the combinatorial distribution of acoustic samples. The x-axis represents the external container materials, and the y-axis indicates the number of audio recordings, broken down by internal fill materials.}
  \label{fig:04_fine_grained_stacked_bar_updated}
\end{figure}

\textbf{Fine-grained Internal States.} 
The core contribution of our dataset lies in its meticulous annotation of internal physical states. First, we introduce distinct fill conditions: liquids (water), granular media (rice), and an empty baseline (hollow), which exhibit fundamentally different damping and added-mass acoustic behaviors. Figure \ref{fig:02_fill_material_pie_updated} presents the proportional distribution of the audio recordings based on these internal states. To ensure our model learns robust cross-modal physical priors, we record a highly balanced combinatorial distribution of these fill materials across all six container types. Figure \ref{fig:04_fine_grained_stacked_bar_updated} illustrates this comprehensive cross-distribution, where the x-axis represents the external materials and the y-axis details the volume of audio samples. This demonstrates that different internal states are extensively sampled within each external shell category. Combined with recordings across continuous fill levels and distinct striker types, this rigorous protocol ensures that FillGauss learns a continuous physical mapping between 3D geometry, internal semantics, and impact acoustics.

\end{document}